\documentclass[twocolumn,floatfix,tighten]{aastex631}
\usepackage{amsmath} 
\submitjournal{ApJ}

\begin{document}
\title{Stellar $\beta^{-}$-decay rate of $^{63}$Ni and its impact on the $\emph{s}$-process nucleosynthesis in massive stars}

\correspondingauthor{Bao-Hua Sun}
\email{bhsun@buaa.edu.cn}

\correspondingauthor{Toshitaka Kajino}
\email{kajino@buaa.edu.cn}

\author{Xin-Xu Wang}
\affiliation{School of Physics, Beihang University, Beijing 100191, P. R. China}

\author{Bao-Hua Sun}
\affiliation{School of Physics, Beihang University, Beijing 100191, P. R. China}

\author{Toshitaka Kajino}
\affiliation{School of Physics, Peng Huanwu Collaborative Center for Research and Education, International Research Center for Big-Bang Cosmology and Element Genesis, Beihang University, Beijing 100191, P. R. China}
\affiliation{National Astronomical Observatory of Japan, 2-21-1 Osawa, Mitaka, Tokyo 181-8588, Japan}
\affiliation{Graduate School of Science, The University of Tokyo, 7-3-1 Hongo, Bunkyo-ku, Tokyo 113-0033, Japan}

\author{Zhen-Yu He}
\affiliation{School of Physics, Peng Huanwu Collaborative Center for Research and Education, International Research Center for Big-Bang Cosmology and Element Genesis, Beihang University, Beijing 100191, P. R. China}

\author{Toshio Suzuki}
\affiliation{Department of Physics, College of Humanities and Sciences, Nihon University, 3-25-40 Sakurajosui, Setagaya-ku, Tokyo 156-8550, Japan}
\affiliation{NAT Research Center, NAT Corporation, 3129-45 Hibara Muramatsu, Tokai-mura, Naka-gun, Ibaraki 319-1112, Japan}

\author{Dong-Liang Fang}
\affiliation{Institute of Modern Physics, Chinese Academy of Sciences, Lanzhou 730000, P. R. China}

\author{Zhong-Ming Niu}
\affiliation{School of Physics and Optoelectronic Engineering, Anhui University, Hefei 230601, P. R. China.}



\begin{abstract}

The $\beta^{-}$-decay rate of $^{63}$Ni, an important branching point, affects the subsequent nucleosynthesis in the weak component of the slow-neutron capture process (weak $s$-process).
To evaluate the impact of the uncertainties of stellar lifetime of $^{63}$Ni on abundances, we calculate the contribution to $\beta^{-}$-decay rates from its excited states using the large-scale shell model with various interactions and also explore the atomic effects in the highly ionized plasma.
In the core He burning stage and the shell C burning stage of massive stars, our new rates can be larger than those from Takahashi and Yokoi(1987) by up to a factor of 4 and 6, respectively.
We evaluate the impact of the stellar decay rates of $^{63}$Ni on 
the nucleosynthesis of $A=60\sim90$
in a star with an initial mass of 25 $M_{\bigodot}$ and solar metalicity. 
We find that the new rates can lead to the abundance changes of $^{64}$Ni, $^{63}$Cu, $^{65}$Cu, $^{64}$Zn, $^{66}$Zn, $^{67}$Zn, and $^{68}$Zn by up to $18\%$, $14\%$, $7\%$, $98\%$, $16\%$, $15\%$, and $13\%$,  respectively, after the shell C burning stage at the Lagrangian mass coordinate $M_{r}=2M_{\bigodot}$. 
The enhancement of the decay rate of $^{63}$Ni increases the weak $s$-process efficiency of nuclei after $^{65}$Cu. 
\end{abstract}

\
\
\section{Introduction} \label{sec:intro}
 The weak component of the slow-neutron capture process (weak $s$-process)
 in massive stars ($M_{\rm{initial}}>(8\sim10)$ $M_{\bigodot}$) can be the main source of $s$-nuclei with $A=60\sim90$ \citep{1972ApJ...173..637P,1989RPPh...52..945K,1990A&A...234..211P,pignatari2008,Pignatari_2010}. 
It is widely believed to occur in two stages: the core He burning and the shell C burning.
Most of the $s$ yields from the core He burning are modified by the shell C burning.
Many observational data and theoretical calculations suggest that about $90\%$ of the solar Cu originates from the weak $s$-process and about $5\%$ from 
the main component of the $s$-process (main $s$-process).
Concerning Zn, $50\%$ of solar Zn can be synthesized during the weak $s$-process \citep[and references therein]{RevModPhys.83.157}. 

However, there are still several open questions concerning the origin of Cu and Zn.
Firstly, to reproduce the solar Cu isotopes, \cite{Pignatari_2010} pointed out that the $^{63}$Ni($n$,$\gamma$)$^{64}$Ni cross section should be half of the theoretical value from the NON-SMOKER code~\citep{Dillmann2010}. 
In contrast, the experimental Maxwellian-averaged cross section of $^{63}$Ni($n$,$\gamma$)$^{64}$Ni is larger than the theoretical value by a factor of 2, leading to a decrease in the abundances of $^{63}$Cu and $^{64}$Zn, while the abundances of $^{64}$Ni and subsequent heavier $s$-nuclei increase \citep{6263nitof,63nidance}.
Remaining uncertainties due to unresolved nuclear physics (i.e., $\beta$-decay rates) could be the key to solving the problem.
The second question concerns the origin of Zn in extremely metal-poor (EMP) stars.
A formation model of C-rich EMP stars underpredicts the Zn abundance in the EMP star CS 22949-037.
These stars are believed to burn from the mixing of the ejecta of almost metal-free supernovae and EMP interstellar matter \citep{2003Natur.422..871U}.

The yields of Cu and Zn in the weak $s$-process can be modified by the $\beta$-decays or the neutron captures.  
$^{63}$Ni servers as an important branching point and bottleneck of the weak $s$-process path in Ni-Cu-Zn region \citep{RevModPhys.83.157}. 
The competition between $^{63}$Ni($n$,$\gamma$)$^{64}$Ni and $^{63}$Ni($\beta^{-}\bar{\nu}$)$^{63}$Cu can affect the abundances of stable nuclei with $A=63\sim90$.
The known laboratory half-life of $^{63}$Ni is $101.2\pm1.5$ yr \citep{COLLE200860}.
It refers to the decay from the ground state (g.s.) of the neutral $^{63}$Ni atom ($Q_{g.s.}$=67 keV).

However, the $\beta^{-}$-decay rates may be significantly elevated in the stellar environment.
Firstly, several allowed and first non-unique forbidden transitions can be involved in the total decay channel due to the thermal equilibrium between g.s. and low-lying excited states \citep{1989RPPh...52..945K,Nb92,betaFe59,59Febeta,Li_2021}. 
Secondly, many $^{63}$Ni atoms are highly ionized in the high-temperature and high-density stellar plasma. 
A new decay channel known as bound-state $\beta^{-}$-decay can open in such a scenario.
The $\beta^-$ particle can be emitted not only into free space but also can occupy open electron orbits \citep{TAKAHASHI1983578,1989RPPh...52..945K,firstbound,187Re,simBound,Limongi_2000,heavybound}. 

\cite{TAKAHASHI1987375}
systematically calculated the stellar $\beta$-decay rates of heavy nuclides ($26\lesssim Z\lesssim83$, $59\lesssim A\lesssim210$, hereafter refer to TY87).
The decay rates were provided only at $T\lesssim 0.5$ GK, but the temperature of weak $s$-process can exceed 1 GK.
In recent decades, the shell model (SM) with a fully diagonalized effective Hamiltonian in space has been successfully applied to solve weak interaction rates of $sd$-shell nuclei \citep{10,25,26,29} and $pf$-shell nuclei \citep{27,28,29}.
Langanke and Martínez-Pinedo calculated the weak interaction rates (both electron capture and $\beta$-decay) of atomic nuclei with $45\lesssim A\lesssim65$ in a range of $0.01\lesssim T\lesssim100$ GK and $10\lesssim \rho Y_{e}\lesssim10^{11}$ g cm$^{-3}$ \citep{28}.
Gupta et al. explored the allowed $\beta^{-}$ decay of fully ionized atoms with $A=60\sim80$ using the realistic nuclear shell model in the stellar environment \citep{PhysRevC.108.015805}.
However, the $\beta^{-}$-decay rates of $^{63}$Ni obtained from SM calculations are theoretically inefficiently evaluated.

In this paper, we calculate the Gamow-Teller (GT) transition strengths $B$(GT) of the $\beta^{-}$-decays from the two low-lying excited states of $^{63}$Ni at 87.2 and 155.6 keV using the large-scale shell model (section \ref{subsec:excited}) by taking account of the atomic effects from highly ionized atoms (section \ref{subsec:ionized}) to obtain the stellar $\beta^{-}$-decay rates.
We extensively calculate the decay rate up to 1.2 GK which well covers the realistic temperature for the weak $s$-process.
We then study the impact of our improved $\beta^{-}$-decay rates on the weak $s$-process nucleosynthesis using a 25 $M_{\bigodot}$ star model with solar metalicity.  

\section{Concepts of stellar \texorpdfstring{$\beta$}{Lg}-decays} \label{sec:Ther}

In the stellar environment, the time scales of the allowed $\gamma$-transitions between low-lying excited states and g.s. are usually much shorter than the typical time scale of the $s$-process.
This ensures that the system of $\beta$-unstable nuclei is in equilibrium through the entire nucleosynthesis.
The ratios of the population of excited states to that of g.s. can be obtained by the Boltzmann distribution, and the contribution of these $\beta$-decays from excited states will accordingly modify the total decay rate \citep{1989RPPh...52..945K}.
In addition, atoms can be highly ionized under local thermodynamical equilibrium in the stellar environment.
Different ionized states are populated by following the Saha equation.

The stellar $\beta^{-}$-transitions from the initial state X to the final state Y have two modes:

\noindent continuum $\beta^{-}$-decay
\begin{equation} 
^{A}_{Z}{\rm X}^{j+}_{ik} \rightarrow ^{\quad A}_{Z+1}{\rm Y}^{(j+1)+}_{i^{'}k^{'}} + e^{-} + \bar{\nu}\;, \nonumber
\end{equation}

\noindent and bound-state $\beta^{-}$-decay
\begin{equation} 
^{A}_{Z}{\rm X}^{j+}_{ik} \rightarrow _{Z+1}^{\quad A}{\rm Y}^{j+}_{i^{'}k^{'}} + \bar{\nu}\;,\nonumber
\end{equation}
where $i$ and $k$ are the nuclear and atomic states, $j$ the degree of ionization.
It should be noted that the effect of the atomic excited states on the decay rate can be neglected in the present work. We only discuss the allowed $\beta ^{-}$-transitions in the subsequent sections.

Considering all the $\beta^{-}$-transitions from the initial state X in different excited states and ionized states, the total stellar $\beta^{-}$-decay rate $\lambda_{\rm stellar}$ from the initial state $^{A}_{Z}{\rm X}$ can be written as
\begin{equation}\label{15}
\lambda_{\rm stellar} = \sum_{i}[ P_{i}\times \sum_{i^{'}}( I_{\beta^{-}}^{ii^{'}} \times \lambda_{n_{ii^{'}}} )]\;, 
\end{equation}
where $I_{\beta^{-}}$ is introduced as the ratio of the sum of continuum and bound-state decay rates to the corresponding rate $\lambda_{n_{ii^{'}}}$ of the neutral atom for a certain $\beta^{-}$-transition.
$P_{i}$ is the thermally populated probability of each state, 
\begin{equation} \label{16}
P_{i} = \frac{(2J_{i}+1)exp(-E_{i}/\kappa T)}{\sum_{l}(2J_{l}+1)exp(-E_{l}/\kappa T)}\;.
\end{equation}
Here $J_{i}$ is the spin.
The detailed formulas are shown in Appendix \ref{stellar}

\section{Stellar \texorpdfstring{$\beta^{-}$}{Lg}-decay rate of \texorpdfstring{$^{63}$}{Lg}N\lowercase{i}} \label{subsec:63Nidecay}
\subsection{\texorpdfstring{$\beta^{-}$}{Lg}-decay Rates from Excited States} \label{subsec:excited}

\begin{figure}[ht!]
\centering
\includegraphics[width=1.0\linewidth]{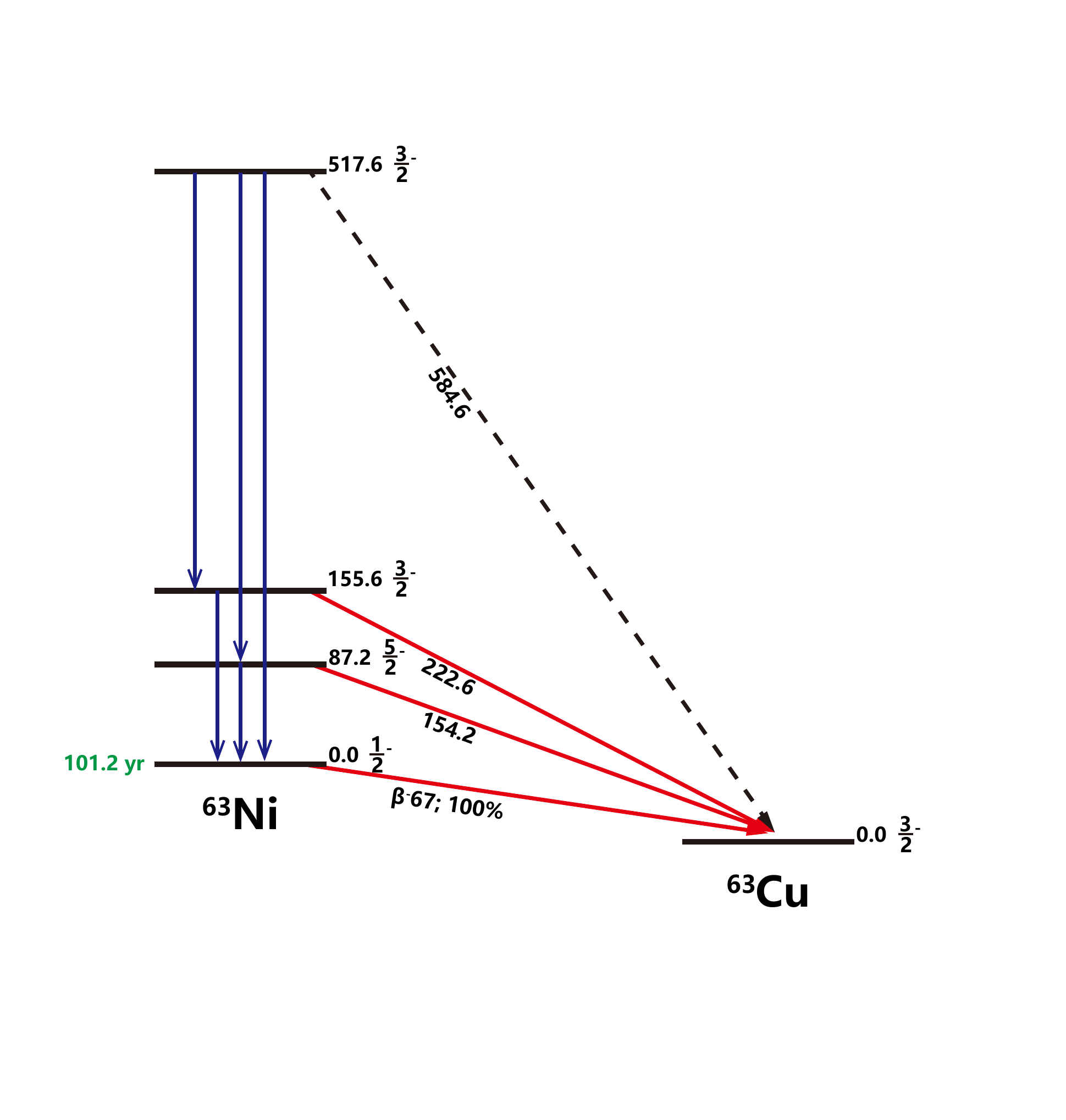}
\caption{Stellar $\beta^{-}$-decay of $\rm{^{63}Ni}$. All the $\beta$-transitions labeled are allowed transitions. 
The red lines represent the dominant ones in the weak $s$-process. 
Data are from \citealt{RUNTE1985237,COLLE200860}.}  
\label{fig1}
\end{figure}

In this subsection, we focus on the decay rates affected by the temperature and take the $I_{\beta^{-}}$ to unity. 
The $\beta^{-}$-decay scheme of $\rm{^{63}Ni}$ is shown in Figure \ref{fig1}. 
The $\gamma$-transition life from the first excited state (87.2 keV) to g.s. is $1.61\pm0.07$ ${\mu}$s \citep{Roodbergen1975TransitionPI}, which is much shorter than the time scale of shell C burning $\tau\sim0.39$ yr \citep{Limongi_2000}. 
Therefore, the system is in thermal equilibrium during the weak $s$-process in the epoch of shell C burning.
The ratios of the population probabilities of excited states at 87.2 keV, 155.6 keV, and 517.6 keV relative to g.s. are 1.09, $3.29\times10^{-1}$, and $4.93\times10^{-3}$ at 1 GK, which is the typical temperature of shell C burning.
Thus, the $\beta^{-}$-transitions from g.s., and the two low-lying excited states at 87.2 and 155.6 keV dominate in the weak $s$-process.

We take the experimental rate for g.s. $\rightarrow$ g.s. transition.
The $\beta^{-}$-decay rates from the two low-lying excited states have not been measured experimentally. 
TY87 showed the log$ft$ values of 6.5 and 5.5 for these states.
We calculate the decay rates from excited states in the large-scale shell-model in the $pf$-shell.
The wave functions of g.s. and two low-lying excited states have major components which occupy the $p_{1/2}$, $f_{5/2}$, and $p_{3/2}$ orbitals. 
The ground state of $^{63}$Cu has $p_{3/2}$ orbital configuration.

To calculate the decay rates from excited states, we adopt three different interactions (fpd6pn, GXPF1J and jun45) which have been successfully applied in the mass range near $A=63$. 
The fpd6 interaction was obtained by fitting energy levels of $fp$-shell nuclei at $A=41\sim49$ \citep{RICHTER1991325} and then was extended to $A=41\sim66$ \citep{VANDERMERWE1994173}. 
The GXPF1J interaction \citep{Honma2005} was developed from GXPF1 \citep{Honma2002,Honma2004} by fitting the energy position of 1$^+$ state in $^{48}$Ca at the experimental energy, $E_x =10.23$ MeV.
This interaction is used to successfully reproduce the experimental GT strength distribution in $^{56}$Ni,  especially its two-peak structure \citep{PhysRevC.79.061603} and to study The GT$_{+}$ strength and electron capture rates of Ni isotopes \citep{PhysRevC.83.044619}. 
The jun45 interaction was derived by fitting experimental energy data from 69 nuclei in $A=63\sim96$ \citep{PhysRevC.80.064323} and used to study the single-particle and collective excitations of $^{63}$Ni \citep{PhysRevC.88.054314}.

\begin{figure}[ht!]
\centering
\includegraphics[width=1.0\linewidth]{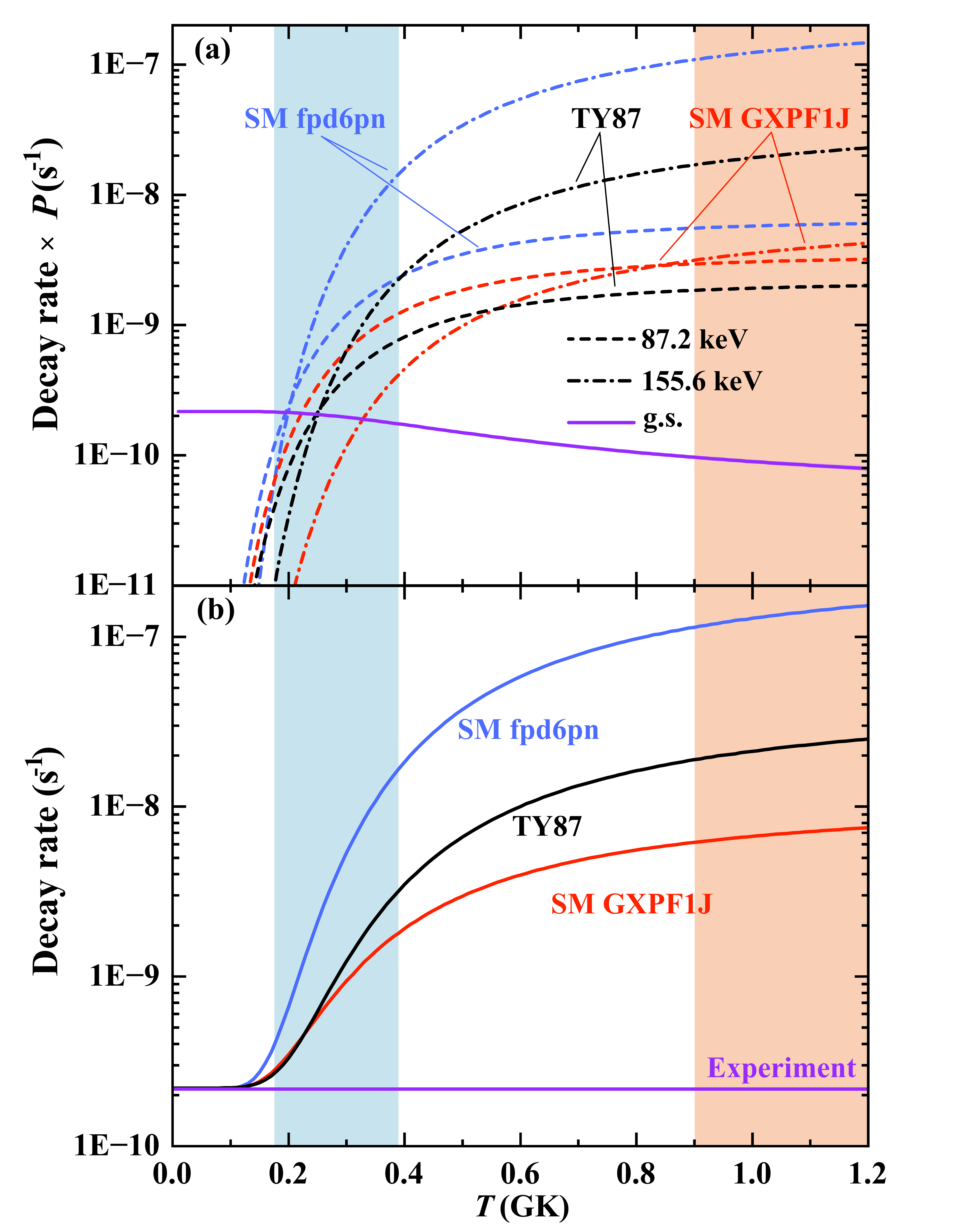}
\caption{Decay rates from the experiment (purple), TY87 (black), SM fpd6pn (blue) and SM GXPF1J (red) as a function of temperature at $n_{e}=0$.
The light-blue and orange bands represent the temperature ranges in the core He burning and the shell C burning stages, respectively.
(a) Decay rates from the g.s. (solid), 87.2 keV state (dashed), and 155.6 keV state (dashed-dotted) of $^{63}$Ni.
These rates include the populated probabilities of the states.
(b) Total $\beta^{-}$-decay rates.  }
\label{fig2}
\end{figure}

Table~\ref{tab:bgt} presents the $B$(GT) results of our shell model calculations based on above three different interactions, together with the published results using the fpd6npn interaction~\citep{PhysRevC.108.015805}.  
For GT transitions, $B$(GT) links to the half-life by:
\begin{equation}\label{3}
(\frac{g_{A}}{g_{V}})^{2}B({\rm GT}) = \frac{K/g^{2}_{V}}{ft} \;,
\end{equation}
where ${g_{A}}/{g_{V}}=-1.2694 \pm 0.0028$ is the ratio of axial to vector coupling constants, and $K/g^{2}_{V}$ is $6143\pm2$ s \citep{PhysRevC.86.015809}.
At the temperature of shell C burning, the most important transition is $^{63}$Ni (3/2$^{-}$, 155.6 keV) $\rightarrow$ $^{63}$Cu (3/2$^{-}$, g.s.). The highest value for this transition is from SM fpd6pn (Table \ref{tab:bgt}), and the lowest is from SM jun45. 
For the transition of $^{63}$Ni (5/2$^{-}$, 87.2 keV) $\rightarrow$ $^{63}$Cu (3/2$^{-}$, g.s.), the TY87 gives the lowest values, while SM jun45 predicts the highest.
Taking account of the relative population probabilities of 1.09 for the 87.2 keV state and 0.329 for the 155.6 keV state, the highest total decay rate is from SM fpd6pn, and the lowest one is from SM GXPF1J at 1 GK. 
We, therefore, show these SM decay rates and compare them with the rate from the TY87 in Figure~\ref{fig2}.

\begin{deluxetable}{ccccc}
\tablecaption{$B$(GT) values for allowed $\beta^{-}$-transitions from the first two low-lying excited states of $^{63}$Ni to the ground state of $^{63}$Cu \label{tab:bgt}}
\tablehead{
\colhead{Model} & \colhead{87.2 keV} & \colhead{155.6 keV} & \colhead{Reference}
}
\startdata
TY87 & $1.21\times10^{-3}$ & $1.21\times10^{-2}$ & TY87 \\
SM fpd6pn &$3.56\times10^{-3}$ & $7.78\times10^{-2}$ & This work \\
SM GXPF1J&$1.90\times10^{-3}$ & $2.25\times10^{-3}$ & This work \\
SM jun45 &$4.48\times10^{-2}$ & $8.73\times10^{-4}$ & This work \\
SM fpd6npn &$6.83\times10^{-3}$ & $1.81\times10^{-2}$ & \cite{PhysRevC.108.015805} \\
\enddata
\tablecomments{The experimental $B$(GT) for g.s. $\rightarrow$ g.s. $\beta^{-}$-transition is $(7.6\pm0.8)\times10^{-4}$.}
\end{deluxetable}

Figure~\ref{fig2}(a) shows the decay rates from three states in the above cases.
The ranges of the core He burning (light-blue) and the shell C burning (orange) are from \citealt{He_2020}.
The g.s. $\rightarrow$ g.s. transition dominates at $T\lesssim0.2$ GK.
The total decay rate is mainly influenced by the joint contribution of the two excited-state decays during the core He burning.
The 155.6 keV $\rightarrow$ g.s. transition rate is much higher than the transition rates of the other two states during the shell C burning, except for the rate from SM GXPF1J.

The total $\beta^{-}$-decay rates are shown in Figure~\ref{fig2}(b).
These rates can be $30\sim600$ times the experimental rate during the shell C burning. 
The decay rates from SM fpd6pn are enhanced by up to about a factor of 4 and 6 ($T_{1/2}\sim0.17$ yr), compared to the rates from the TY87 at $0.3$ and $1$ GK, respectively.
Under the same comparison, the rate from SM GXPF1J is reduced by about a factor of 2 and 3 at the corresponding temperatures, respectively.

\subsection{Continuum and Bound-state \texorpdfstring{$\beta^{-}$}{Lg}-decays in Different Ionized States} \label{subsec:ionized}

\begin{figure}[ht!]
\centering
\includegraphics[width=1.0\linewidth]{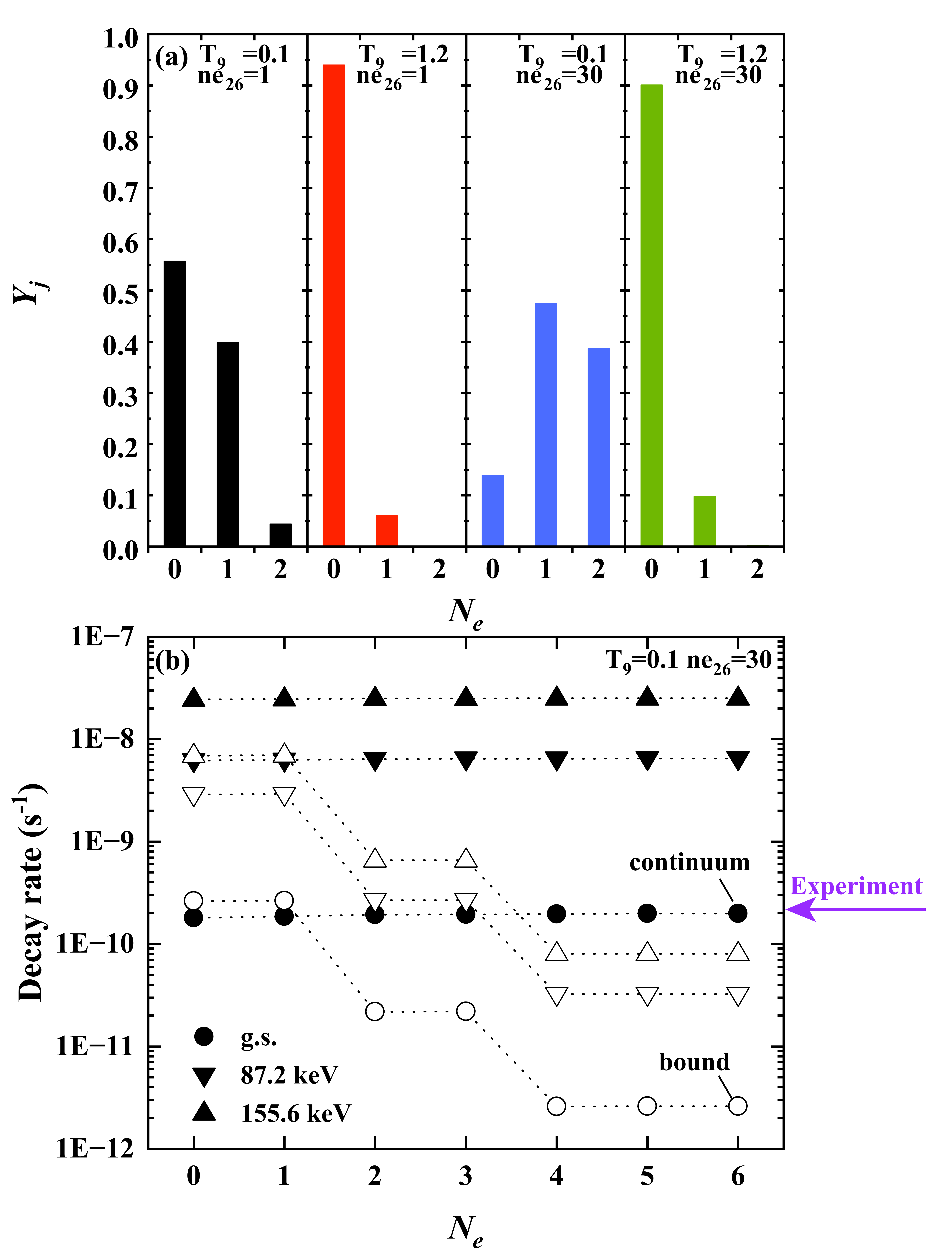}
\caption{(a) Distribution of three main ionized states of $\rm{^{63}Ni}$ in four stellar cases ($T$=0.1 or 1.2 GK, and $n_{e}$=1$\times$10$^{26}$ or 30$\times$10$^{26}$ cm$^{-3}$). 
$N_{e}$ is the number of bound electrons.
$Y_{j}$ is the abundance of ions with $N_{e}$ electrons.
$T_{9}$ is the temperature in the unit of $10^{9}$ K.
$ne_{26}$ is the electron number density in unit of 10$^{26}$ cm$^{-3}$.
(b) Continuum (full symbols) and bound-state (open symbols) $\beta^{-}$-decay rates calculated using the SM GXPF1J interaction for the g.s. and the first two low-lying excited states as a function of $N_{e}$ at $T=0.1$ GK and $n_{e}=3\times10^{27}$ cm$^{-3}$.   
The arrow indicates the experimental rate. }
\label{fig3}
\end{figure}

In stellar environments, most of $^{63}$Ni atoms are in fully ionized, H-like ($1s_{1/2}$), or He-like ($2s_{1/2}$) states as shown in Figure \ref{fig3}(a). As an example, the states are populated with about 14$\%$, 47$\%$, and 39$\%$ at $T=0.1$ GK and $n_{e}=3\times10^{27}$ cm$^{-3}$. The abundances of highly ionized atoms increase with increasing temperature but decrease with increasing electron number density $n_{e}$. The percentage of fully ionized and H-like atoms exceeds 60$\%$ at $T=0.1$ GK and becomes closer to 100$\%$ at $T=1.2$ GK when $n_{e}$ is of the order of $10^{27}$ cm$^{-3}$. 

The decay rates of each nuclear state can be enhanced by adding the new decay channel called bound-state $\beta ^{-}$-decays of charged ions.
Figure \ref{fig3} (b) shows the rates of the continuum and bound-state $\beta ^{-}$-decay predicated using the SM GXPF1J interaction for three states of $^{63}$Ni at $T=0.1$ GK and $n_{e}=3\times10^{27}$.  
The number of bound electrons of $^{63}$Ni only has a minor impact on the continuum decay rates but has a significant impact on the bound-state decay rates.
Filling electron orbital induces a sudden drop in the decay rates.
For the g.s. $\rightarrow$ g.s. transition, the bound-state decay rates in the fully ionized and H-like states are comparable to the continuum decay rates, because the decay energy (67 keV) and the K shell binding of the daughter nuclei (11.57 keV, \citealt{cuatom}) are in the same order of magnitude.
For the decay from two excited states, the bound-state rates are lower than the continuum rates.
Despite this, including bound-state $\beta^{-}$-decays results in an increase of at least $20\%$ in the total decay rate of each excited state at this temperature and density (see Table \ref{tab:ionized}).

\subsection{Total \texorpdfstring{$\beta^{-}$}{Lg}-decay Rates} \label{subsec:total}

\begin{figure}[ht!]
\centering
\includegraphics[width=1.0 \linewidth]{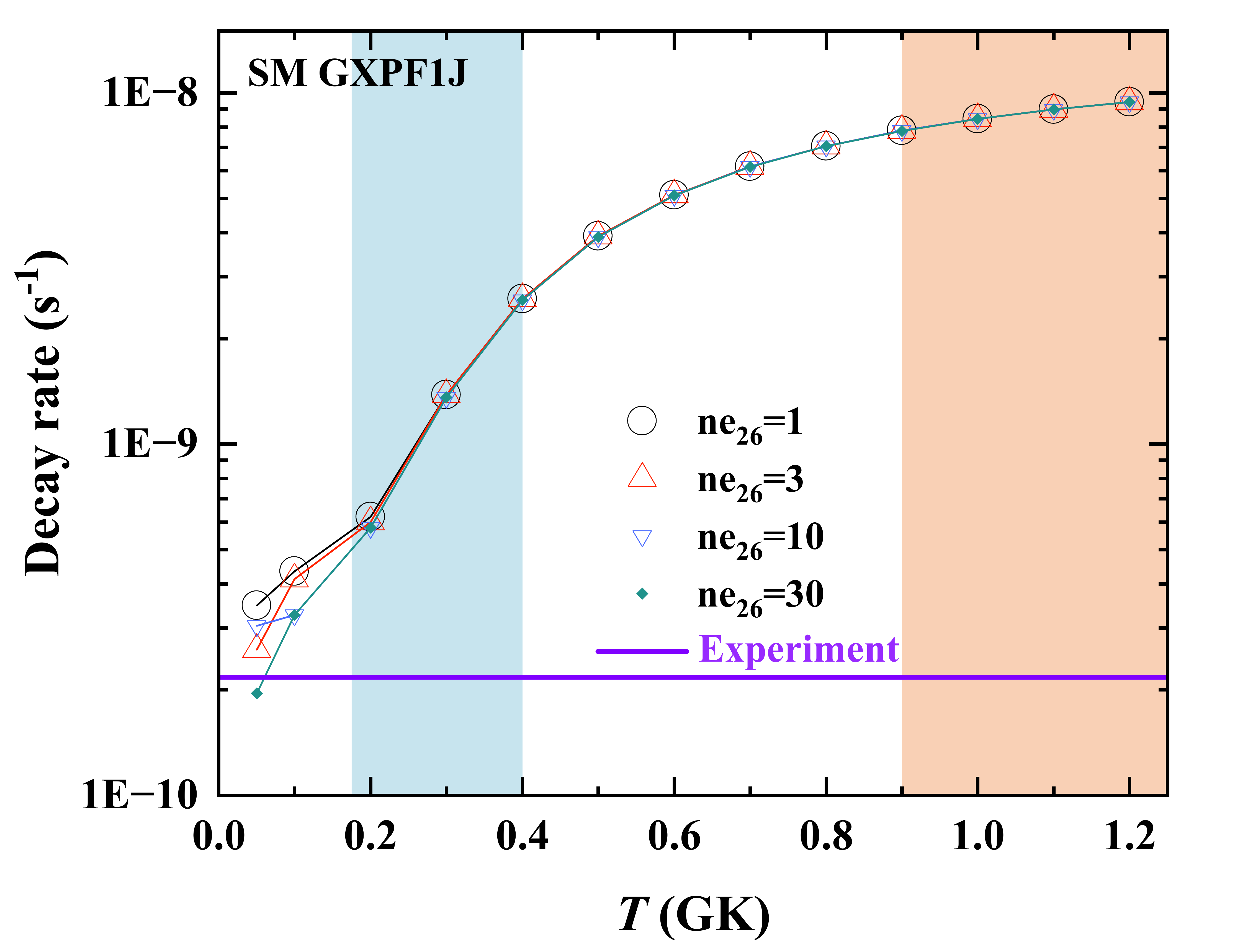}
\caption{Stellar $\beta^{-}$-decay rates of $^{63}$Ni using the SM GXPF1J interaction as a function of temperature in four electron number densities, $n_{e}=1,3,10,30\times10^{26}$ cm$^{-3}$. 
The laboratory rate is shown for comparison.}
\label{fig4}
\end{figure}

The total stellar decay (Equation~\ref{15}) gets fast with increasing temperature but slows with increasing electron density by taking account of the effect from excited states and highly ionized atoms.
Figure~\ref{fig4} shows the stellar $\beta^{-}$-decay rates of $^{63}$Ni using the SM GXPF1J interaction in four electron number densities.
The adopted $I_{\beta^{-}}$ factors are listed in Appendix \ref{SF}.
Compared to the laboratory value, the decay as a whole can speed up by a factor of 2 at $T=0.1$ GK and $n_{e}=10^{26}$ cm$^{-3}$, primarily due to the contribution of bound-state $\beta^{-}$-decays. 
The decay rate can be at least enhanced by a factor of 6 in $T\gtrsim0.3$ GK and 40 in $T\gtrsim1$ GK.
We further calculate the stellar $\beta^{-}$-decay rates from SM fpd6pn and extend TY87 to the present temperature and $n_{e}$ region.

\section{Impact of stellar \texorpdfstring{$\beta^{-}$}{Lg}-decay of \texorpdfstring{$^{63}$}{Lg}Ni on \texorpdfstring{$s$}{Lg}-process in massive stars} \label{sec:effect}

The $\beta$-decay branching of $^{63}$Ni can be one of the solutions to the origin of Cu and Zn. 
In this subsection, we explore its impact on the weak $s$-process nucleosynthesis.
The $\beta^{-}$-decay rate from the TY87 is taken as a reference and compared with the results from SM fpd6pn and SM GXPF1J.
We use a multi-zone post-process nucleosynthesis code for a 25$M_{\bigodot}$ star~\citep{He_2020} with solar metalicity.
Some nuclear physics inputs are updated (see Table \ref{tab:list}).
The stellar center at $M_{r}=0M_{\bigodot}$ and $M_{r}=2M_{\bigodot}$ are chosen to discuss the weak $s$-process nucleosynthesis in the core He and shell C burning stages, respectively.

\subsection{Core He Burning} \label{subsec:core}

\begin{figure}[ht!] 
\centering
\includegraphics[width=1.0\linewidth]{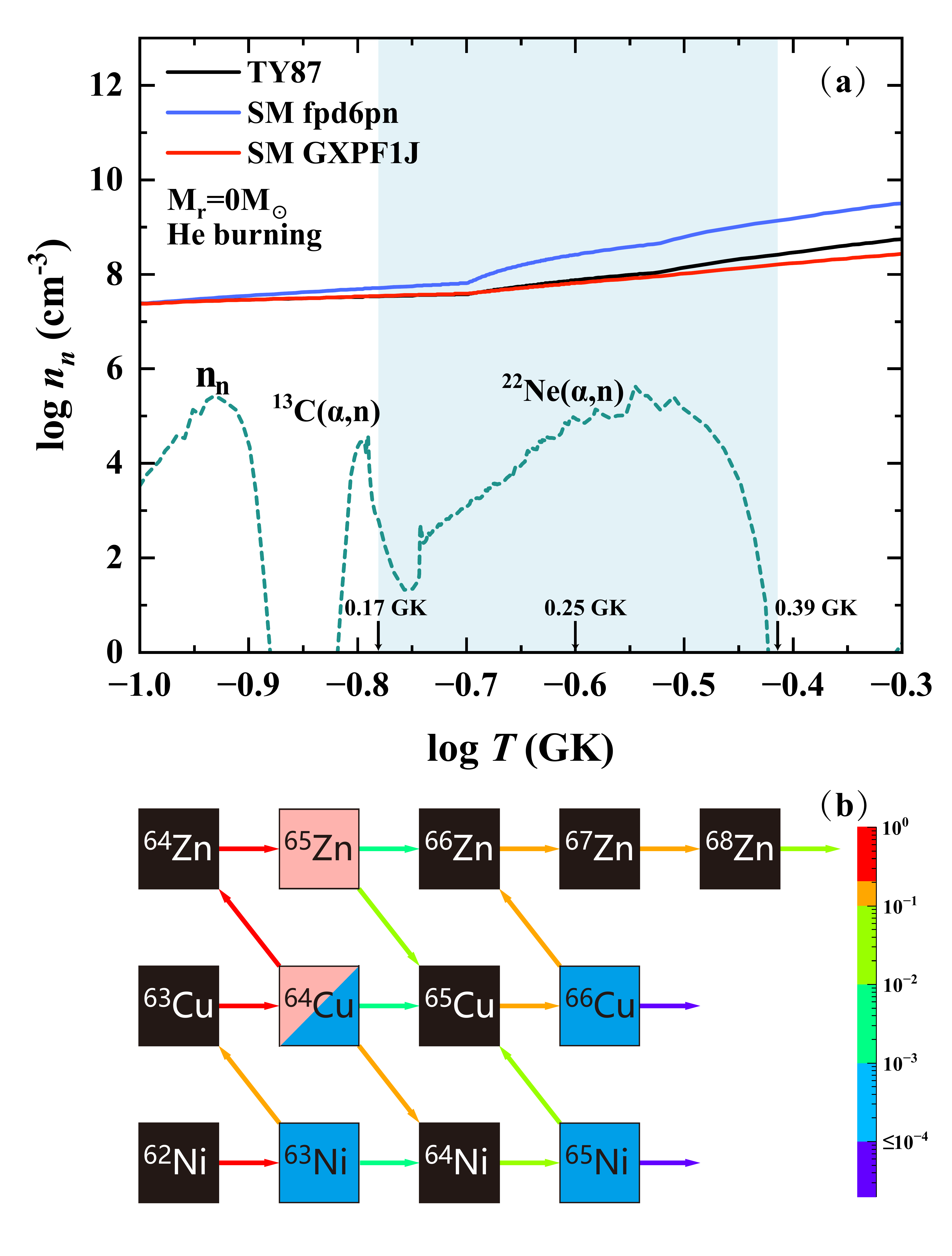}
\caption{(a) Balanced neutron densities (solid lines) at different temperatures for $^{63}$Ni at $M_{r}$=0$M_{\bigodot}$ in a $25M_{\bigodot}$ star with solar metalicity, using the 
decay rates from the TY87 (black), SM fpd6pn (blue), and SM GXPF1J (red). 
For comparison, the green dashed line shows the central neutron number density. 
The light-blue range represents the condition in the core He burning. 
(b) Net mass flow in Ni-Cu-Zn region during the core He burning. The rate of $^{63}$Ni($\beta^{-}\bar{\nu}$)$^{63}$Cu is from the TY87. 
The flow has been scaled by the $^{62}$Ni($n$,$\gamma$) reaction flow.} 
\label{fig7}
\end{figure}

The core He burning is the following stage of the H burning as H is depleted at the center $M_{r}$=0$M_{\bigodot}$.
As shown in Figure~\ref{fig7}(a), the core He burning starts at around 0.17 GK and then completes at 0.39 GK when the mass fraction of $^{4}$He is less than $10^{-8}$.  
$^{22}$Ne($\alpha$,n)$^{25}$Mg as the primary neutron source gradually increases the neutron number density $n_{n}$ and will eventually trigger the weak $s$-process at $T\gtrsim0.25$ GK.
The maximum neutron density is less than $10^{6}$ cm$^{-3}$.

For $^{63}$Ni, we define the balanced neutron density, where the relevant neutron capture rate equals to the decay rate. 
The balanced neutron densities vs. temperatures curves are shown in Figure~\ref{fig7}(a) when using three decay rates.
The balanced neutron densities, typically over $10^7$ cm$^{-3}$, are higher by one order of magnitude than the central neutron density.
Therefore, the $\beta^{-}$-decay always dominates at $^{63}$Ni, as shown in Figure~\ref{fig7}(b).
This decay serves as a key for the subsequent nucleosynthesis of $^{64}$Ni, $^{63,65}$Cu, and $^{64,66-68}$Zn.

\begin{figure}[ht!]
\centering
\includegraphics[width=1.0\linewidth]{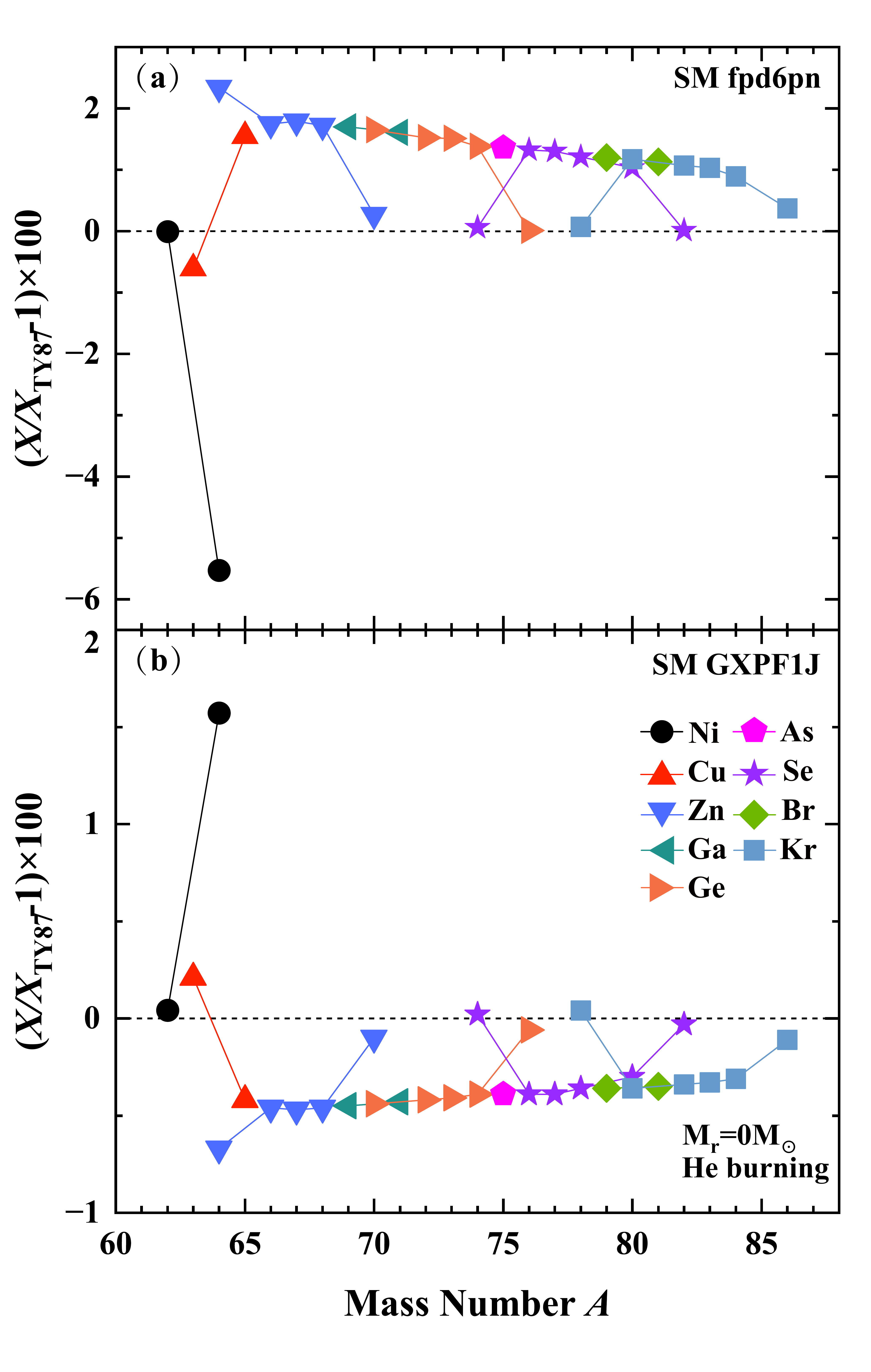}
\caption{Relative difference in mass fractions of nuclei with $\emph{Z}$ = 28 to 36 using the decay rates from (a) SM fpd6pn and (b) SM GXPF1J to those from the TY87, at $M_{r}=0M_{\bigodot}$ after the core He burning.}
\label{fig8}
\end{figure}

The impact of the new SM rates on stable nuclei with $\emph{Z}$ = 28 to 36 is shown in Figure~\ref{fig8} at the end of core He burning.
We calculate the relative changes of the mass fractions to those from the TY87. 
The increased rate of $^{63}$Ni from SM fpd6pn enhances the nucleosynthesis efficiency of subsequent elements
but reduces the net mass flow to $^{64}$Ni. 
As a result, the abundance of $^{64}$Ni decreases by $\gtrsim5\%$ in the SM fpd6pn case and increases by $\gtrsim1\%$ in the SM GXPF1J, relative to the results of the TY87.  
$^{63}$Cu is governed by the final radioactive decay of $^{63}$Ni, so its abundance decreases with increasing decay rate.
The abundances of the other stable nuclei vary within $3\%$ in the SM fpd6pn and $1\%$ in the SM GXPF1J.

\subsection{Shell C Burning} \label{subsec:shell}

\begin{figure}[ht!]
\centering
\includegraphics[width=1.0 \linewidth]{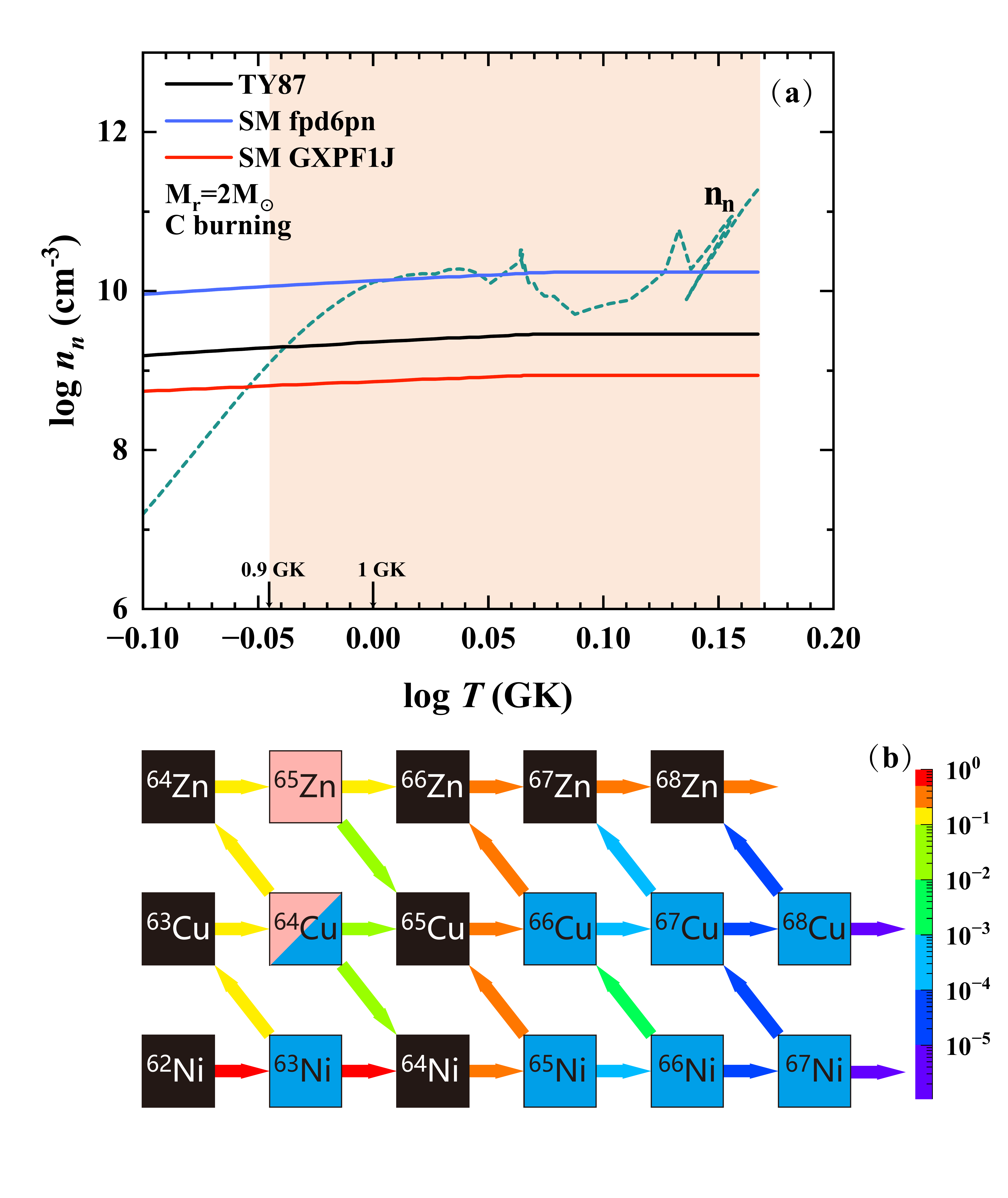}
\caption{Same as Figure~\ref{fig7}, but at $M_{r}$=2$M_{\bigodot}$.
The orange range is the shell C burning. }
\label{fig10}
\end{figure}

After the core He burning, the temperature and density of the core rise to the onset of C burning. 
The C burning continues in the shells after the depletion of the central carbon.
The shell C burning is triggered at around 0.9 GK at $M_{r}$=2$M_{\bigodot}$, as shown in Figure~\ref{fig10}(a).
The main neutron source is the $^{22}$Ne($\alpha$, n) reaction. $^{22}$Ne survives from the previous core He burning, and $\alpha$-particles are from $^{12}$C($^{12}$C,$\alpha$)$^{20}$Ne.
The second round of the $s$-process starts around 1 GK, when nearly half of $^{12}$C is consumed.       
The neutron density $n_{n}$ increases from $10^{9}$ to $10^{11}$ cm$^{-3}$ at $M_{r}$=2$M_{\bigodot}$. 

The balanced neutron densities for the $^{63}$Ni($\beta^{-}\bar{\nu}$)$^{63}$Cu rates from three models are also shown in Figure \ref{fig10}(a). 
The decay rates from the TY87 and SM GXPF1J are lower than the neutron capture rate from the beginning of shell C burning.  
Therefore, more than $80\%$ of $^{63}$Ni is consumed by the $^{63}$Ni($n$,$\gamma$)$^{64}$Ni. This is indicated in Figure \ref{fig10}(b).
The decay rate from SM fpd6pn is comparable to the neutron capture rate at $T\gtrsim1$ GK, but the branching ratio of the $\beta$-decay out of total mass flow through $^{63}$Ni, $\beta^{-}/(\beta^{-}+(n,\gamma))$, decreases to $\sim0.02$ in the late epoch of C burning. 
The net mass flows significantly increase compared to those in the He burning stage. 
The nuclei are mainly produced or consumed by neutron capture reactions due to the high neutron density, and the weak $s$-process path deviates from the $\beta$-stability line by 2 to 3 mass units during the shell C burning.

\begin{figure}[ht!]
\centering
\includegraphics[width=1.0 \linewidth]{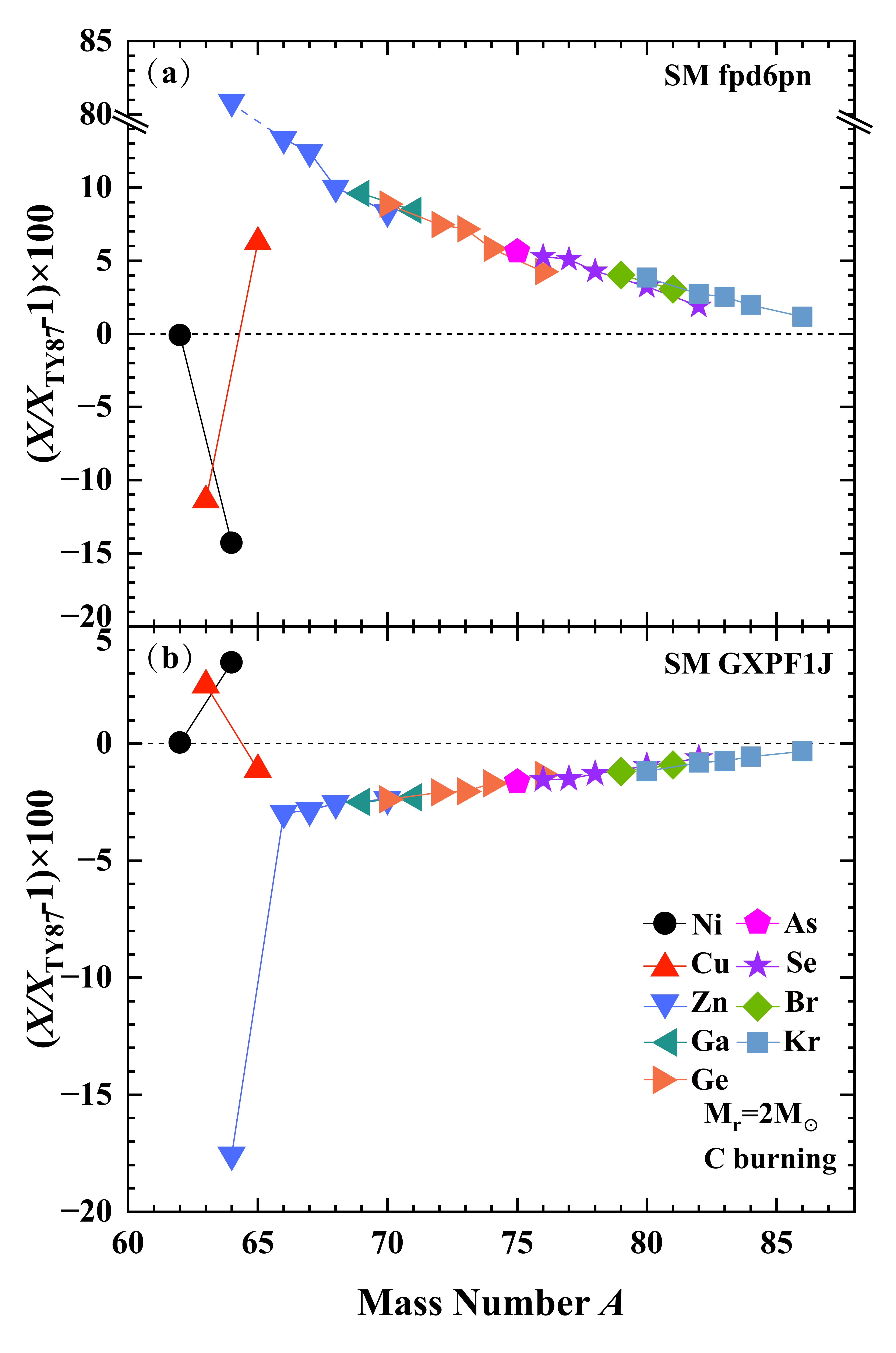}
\caption{Same as Figure \ref{fig8}, but at $M_{r}$=2$M_{\bigodot}$ after the end of shell C burning.}
\label{fig12}
\end{figure}

Figures \ref{fig12} (a)-(b) are the same as Figure \ref{fig8} but at $M_{r}$=2$M_{\bigodot}$ after the end of shell C burning.
The decay rate of $^{63}$Ni significantly affects the abundances of $^{64}$Ni, Cu, and Zn, but its impact is gradually weaker on the abundances from Zn to Kr.
In the SM fpd6pn case, the abundances of $^{64}$Ni and $^{63}$Cu decrease by $14\%$ and $11\%$, and $^{65}$Cu, $^{64, 66-68}$Zn increase by $6\%$, $80\%$, $13\%$, $12\%$, and $10\%$, respectively, compared to those in the TY87.
In general, the mass fraction of Cu decreases by $4\%$, and Zn increases by $15\%$.
In the SM GXPF1J case, $^{64}$Zn can decrease by $17\%$, and the effect is less than $4\%$ in other stable nuclei. 
The mass fraction of Zn decreases by $4\%$, and the relative difference of Cu is only $1\%$.
$^{64}$Zn is the most sensitive nuclide, but its overproduction factor is less than 10 in this stage. 

To fully explore the uncertainties of the stellar decay rate of $^{63}$Ni on predicted abundances, we vary the decay rates of excited states from the TY87 by a factor of 10 (TY87$\times$10 and TY87$\times$0.1).  
This can lead to the abundance changes of $^{64}$Ni, $^{63}$Cu, $^{65}$Cu, $^{64}$Zn, $^{66}$Zn, $^{67}$Zn, and $^{68}$Zn by $25\%$, $19\%$, $11\%$, $189\%$, $25\%$, $23\%$, and $19\%$,  respectively, at the end of shell C burning at $M_{r}=2M_{\bigodot}$.

For a 25 $M_{\bigodot}$ star, the $s$ yields above the Lagrangian mass coordinate $M_{r}=3M_{\bigodot}$ can remain after the explosive nucleosynthesis and be ejected \citep{Pignatari_2010}.
We calculate the total mass fractions of Cu and Zn integrated over $3M_{\bigodot} \lesssim M_{r} \lesssim 25M_{\bigodot}$ before the onset of core Si burning, as shown in Table \ref{tab:yeild}.
The mass fraction of Cu is not sensitive to the decay rates of $^{63}$Ni.
Zn can be enhanced by $3\%$ with the increased decay rate in the SM fpd6pn case, compared to the results from the TY87. 
There can be a difference by $7\%$ in Zn with the TY87 rate multiplied or divided by factor 10.
However, the subsequent burning stages after the core Si burning can modify the products \citep{Nomoto2023}.  
This could enhance the impact of $\beta^{-}-$decay rates in synthesizing heavier nuclei.
\begin{deluxetable}{ccccc}
\tablecaption{Mass fractions of Cu and Zn above $M_{r}$=3$M_{\bigodot}$ before the onset of core Si burning in a $25M_{\bigodot}$ star with solar metalicity \label{tab:yeild}}
\tablehead{
\colhead{Model} & \colhead{Cu} & \colhead{Zn} }
\startdata
TY87 & $2.12\times10^{-4}$ & $2.50\times10^{-4}$  \\
SM fpd6pn & $2.12\times10^{-4}$ & $2.57\times10^{-4}$  \\
SM GXPF1J& $2.12\times10^{-4}$ & $2.48\times10^{-4}$  \\
TY87$\times$10 & $2.12\times10^{-4}$ & $2.60\times10^{-4}$  \\
TY87$\times$0.1 & $2.10\times10^{-4}$ & $2.42\times10^{-4}$  \\
\enddata
\end{deluxetable}

\section{Summary and future prospect} \label{sec:end}

We investigated the stellar $\beta^{-}$-decay rate of $^{63}$Ni and its impact on the weak $s$-process.
We calculated the GT transition strengths from the two low-lying excited states (87.2 and 155.6 keV) of $^{63}$Ni to $^{63}$Cu by the large-scale shell model. 
The $\beta^{-}$ decays from the excited states dominate the total decay rate in the two stages of weak $s$-process: the core He burning and the shell C burning.
We also utilized the Takahashi $\&$ Yokoi's method to calculate the effect from heavy charged ions, including bound-state $\beta^{-}$-decays in the stellar environment. 
The new stellar $\beta^{-}$-decay rate from SM fpd6pn is up to a factor of 6 higher than the TY87, and the rate from GXPF1J is a factor of 3 lower than the TY87 in the shell C burning stage in massive stars.

The new decay rates have been applied to a 1-D multi-zone nucleosynthesis calculation in a 25 $M_{\bigodot}$ star with solar metalicity.
The decay rate from SM fpd6pn enhances the nucleosynthesis of $^{65}$Cu and the following $s$-nuclei with $A\lesssim90$, and causes the decrease of the abundances of $^{64}$Ni and $^{63}$Cu, compared with the result in the TY87.
The lower decay rate from GXPF1J gives the inverse result.
After the core He burning, the abundances of $s$-nuclei are slightly affected by the new rates.
The impact of new rates is remarkable during the shell C burning.
$^{64}$Zn is the nuclide which shows the strongest sensitivity, and its abundances can vary by $98\%$ at $M_{r}$=2$M_{\bigodot}$ after the end of shell C burning among new rates calculated in several shell models.
The abundances of $^{64}$Ni, $^{63}$Cu, $^{65}$Cu, and $^{66-68}$Zn can vary by more than 7$\%$.
A dedicated experiment to study the stellar lifetime of $^{63}$Ni has been approved \citep{Sun}, via  $^{63}$Cu($d,^{2}$He)$^{63}$Ni reaction.

\section{Acknowledgment}
\begin{acknowledgments}

We gratefully acknowledge Motohiko Kusakabe for his in-depth discussions throughout this work.
We thank Michio Honma for the information on the shell-model calculations with the GXPF1J.
We also thank Kohji Takahashi for his help on the calculation of bound-state $\beta^{-}$-decays, and Wenyu Xin for his advice on the evolution of massive stars.
This work was partly supported by the National Natural Science Foundation of China (No. 12325506, 12335009, 11961141004, 12375109, and 12435010), the National Key R$\&$D Program of China (2022YFA1602401), and the CNSA program D050102.
\end{acknowledgments}

%
\vspace{5mm}
\software{MESA (version r11701 \citealt{Paxton_2011})}


\appendix
\section{\texorpdfstring{$\beta^{-}$}{Lg}-decay of highly charged atoms} \label{stellar}
\subsection{Continuum \texorpdfstring{$\beta^{-}$}{Lg}-decay}\label{subsec:con}
The decay rate of the continuum $\beta^{-}$-decay of $^{A}_{Z}{\rm X}^{j+}_{i}$ can be written as
\begin{equation}\label{7}
\lambda_{ii^{'}j}({\rm c.d.}) = \ln2 \frac{f_{ii^{'}j}({\rm c.d.})}{(ft)_{ii^{'}}}\;.
\end{equation}
where $(ft)_{ii^{'}}$ is the usual $ft$ value and $f_{ii^{'}j}(\rm c.d.)$ is the lepton phase volume of the continuum $\beta ^{-}$-decay. 
Here the decay energy $Q_{c}^{ii^{'}j}$ from the initial state $^{A}_{Z}{\rm X}^{j+}_{i}$ to the final state $^{\quad A}_{Z+1}{\rm Y}^{(j+1)+}_{i^{'}}$ is 
\begin{equation}\label{8}
Q_{c}^{ii^{'}j} = [M_{j}({\rm X})_{i} - M_{j+1}({\rm Y})_{i^{'}}]c^{2}.
\end{equation}
$M_{j}({\rm X})_{i}$ is the sum of the ion mass, the nuclear excitation energy, and the ionization potential depression, and can be written as 
\begin{eqnarray} \label{9}
M_{j}({\rm X})_{i}c^{2} = [M_{n}({\rm X})c^{2}-Zm_{e}c^{2}+B_{n}({\rm X})] \qquad\\
+E_{i}({\rm X})+[(Z-j)m_{e}c^{2}-B_{j}({\rm X})] \nonumber \\
+\sum_{j}^{Z-1}\Delta_{m}({\rm X}) \nonumber \;,
\end{eqnarray}
where $B_{n}$ is the total binding energy of atom, $B_{j}$ the binding energy for remaining $(Z-j)$ bound electrons, $m$ the degree of ionization corresponding to the $(Z-m)$-th bound electron. 

In the stellar environment, the binding energy of the $(Z-m)$-th ($0 \lesssim m \lesssim Z$) bound electron for a nucleus with $Z$ protons and $(Z-j)$ ($j \lesssim m$) bound electrons decreases due to the influence of the electron sea and neighboring ions. This is known as ionization potential depression. 
Assuming that the orbital ionization potential is $I_{m}$, the effective ionization potential $\chi_{m}$ is defined to be $I_{m}-\Delta_{m}$, where $\Delta_{m}$ is the continuum depression (see Appendix A in \citealp{TAKAHASHI1983578}).
The case of $\chi_{m}<0$ indicates that the corresponding ionized state does not exist.
If $j$ equals $Z$, the atom is fully ionized.
Takahashi and Yokoi(1983) studied this effect and gave a numerical solution to calculate $\Delta_{m}$. 
They stated that $\Delta_{m}$ is only related to the degree of ionization $j$, temperature $T$, and free electron number density $n_{e}$. 
It should be noted that their numerical solution mainly takes into account heavy nuclei with $A\gtrsim100$.

The decay energy, $Q_{c}^{ii^{'}j}$, can be written as:
\begin{eqnarray} \label{10}
Q_{c}^{ii^{'}j} = [M_{n}({\rm X})-M_{n}({\rm Y})]c^{2} + [B_{n}({\rm X}) \\
-B_{n}({\rm Y})]+ [E_{i}({\rm X})-E_{i^{'}}({\rm Y})] \nonumber \\
-[B_{j}({\rm X})-B_{j+1}({\rm Y})] \nonumber \\
+(\sum_{j}^{Z-1}\Delta_{m}-\sum_{j+1}^{Z}\Delta_{m}) \;. \nonumber
\end{eqnarray}
The $\Delta_{m}$ terms modify the $Q$-values at finite temperature and $n_{e}$.

\subsection{Bound-state \texorpdfstring{$\beta^{-}$}{Lg}-decay}\label{subsec:bound}
The decay rate of the bound-state $\beta ^{-}$-decay can be written as  
\begin{equation}\label{11}
\lambda_{ii^{'}j}{\rm (b.d.)} = \ln2 \frac{f_{ii^{'}j}{\rm (b.d.)}}{(ft)_{ii^{'}}}\;.
\end{equation}
where $f_{ii^{'}j}{\rm (b.d.)}$ is the lepton phase volume of the final $\beta$ particle for the bound-state $\beta ^{-}$-decay.
It can be regarded as the time-mirrored process of the orbital electron capture \citep{Litvinov_2011}.
The decay energy $Q_{b}^{ii^{'}j}$ satisfies the following relation, 
\begin{equation}\label{12}
Q_{b}^{ii^{'}j} = Q_{c}^{ii^{'}j} + B_{shell}({\rm Y})\;,
\end{equation}
where $B_{shell}$ is the binding energy of the electron shell in K, L, M... orbit of the final state Y \citep{PhysRevC.108.015805}. 
Nearly all the decay energy is carried by the monochromatic antineutrino.

\subsection{Total Stellar \texorpdfstring{$\beta^{-}$}{Lg}-decay}\label{subsec:thetotal}
Because there are two lepton phase volumes for the $\beta^-$ particle, the total decay rate $\lambda_{ii^{'}j}$ from the initial state $^{A}_{Z}{\rm X}^{j+}_{i}$ can be written as 
\begin{align}\label{13}
\lambda_{ii^{'}j} &= \ln2 \frac{f_{ii^{'}j}{\rm (c.d.)}+f_{ii^{'}j}{\rm (b.d.)}}{(ft)_{ii^{'}}} \\
&= \lambda_{ii^{'}j}{\rm (c.d.)} + \lambda_{ii^{'}j}{\rm (b.d.)}\;. \nonumber
\end{align} 
$I_{\beta^{-}}$ is the ratio between the total decay rate $\lambda_{ii^{'}j}$ and the corresponding $\beta^{-}$-decay rate $\lambda_{n_{ii^{'}}}$ for a certain $\beta^{-}$-transition of all ionized states:
\begin{align} \label{14}
I_{\beta^{-}}^{ii^{'}}(T,n_{e}) & \equiv \sum_{j}[\frac{\lambda_{ii^{'}j}({\rm total})}{\lambda_{n_{ii^{'}}}} \times Y_{j}] \nonumber \\
& =\sum_{j}[ \frac{\lambda_{ii^{'}j}({\rm c.d.})+\lambda_{ii^{'}j}({\rm b.d.})}{\lambda_{n_{ii^{'}}}}\;\times Y_{j}].
\end{align}
This value is a function of the temperature $T$ and the free electron number density $n_{e}$. 
$Y_{j}$ is the abundance of ions.

\setcounter{table}{0}
\setcounter{figure}{0}
\renewcommand{\thetable}{B\arabic{table}}
\renewcommand{\thefigure}{B\arabic{figure}}
\section{\texorpdfstring{$I_{\beta^{-}}$}{Lg} factors in the stellar environment} \label{SF}
Table \ref{tab:ionized} shows the $I_{\beta^{-}}$ factors in the stellar environment.
We calculate them based on three $\beta^{-}$-transitions from $^{63}$Ni to $^{63}$Cu in the temperature range of $0.05\lesssim T\lesssim1.2$ GK and the electron number density of $1\times10^{26}\lesssim n_{e}\lesssim30\times10^{26}$ cm$^{-3}$. 
At $T=0.05$ GK, $n_{e}=3\times10^{27}$ cm$^{-3}$, the bound-state $\beta^{-}$-decay rate is small because the abundance of Ni$^{26+}$ is $8.7\times10^{-1}$ and the continuum $\beta^{-}$-decay rate is suppressed by the high electron density, so that the factor of g.s. $\rightarrow$ g.s. transition is smaller than 1.

\begin{table}[htbp]
\begin{center}
\caption{$I_{\beta^{-}}$ factors}\label{tab:ionized}
\begin{tabular}{ccccccccccccccc}
\hline
\hline
& $ne_{26}$=1 & 0.05 & 0.1 & 0.2 & 0.3 & 0.4 & 0.5 & 0.6 & 0.7 & 0.8 & 0.9 & 1.0 & 1.1 & 1.2\\
\hline
& g.s. $\rightarrow$ g.s. & 1.6 & 2.0 & 2.1 & 2.1 & 2.1 & 2.1 & 2.1 & 2.1 & 2.1 & 2.1 & 2.1 & 2.1 & 2.1\\
& 87.2 keV$\rightarrow$ g.s. & 1.2 & 1.3 & 1.3 & 1.3 & 1.3 & 1.3 & 1.3 & 1.3 & 1.3 & 1.3 & 1.3 & 1.3 & 1.3 \\
& 155.6 keV$\rightarrow$ g.s. & 1.1 & 1.2 & 1.2 & 1.2 & 1.2 & 1.2 & 1.2 & 1.2 & 1.2 & 1.2 & 1.2 & 1.2 & 1.2 \\
\hline
\hline
& $ne_{26}$=3 & 0.05 & 0.1 & 0.2 & 0.3 & 0.4 & 0.5 & 0.6 & 0.7 & 0.8 & 0.9 & 1.0 & 1.1 & 1.2\\
\hline
& g.s. $\rightarrow$ g.s. & 1.2 & 1.9 & 2.0 & 2.1 & 2.1 & 2.1 & 2.1 & 2.1 & 2.1 & 2.1 & 2.1 & 2.1 & 2.1\\
& 87.2 keV$\rightarrow$ g.s. & 1.1 & 1.3 & 1.3 & 1.3 & 1.3 & 1.3 & 1.3 & 1.3 & 1.3 & 1.3 & 1.3 & 1.3 & 1.3 \\
& 155.6 keV$\rightarrow$ g.s. & 1.0 & 1.2 & 1.2 & 1.2 & 1.2 & 1.2 & 1.2 & 1.2 & 1.2 & 1.2 & 1.2 & 1.2 & 1.2 \\
\hline
\hline
& $ne_{26}$=10 & 0.05 & 0.1 & 0.2 & 0.3 & 0.4 & 0.5 & 0.6 & 0.7 & 0.8 & 0.9 & 1.0 & 1.1 & 1.2\\
\hline
& g.s. $\rightarrow$ g.s. & 1.4 & 1.5 & 1.9 & 2.0 & 2.0 & 2.0 & 2.0 & 2.1 & 2.1 & 2.1 & 2.1 & 2.1 & 2.1\\
& 87.2 keV$\rightarrow$ g.s. & 1.1 & 1.2 & 1.3 & 1.3 & 1.3 & 1.3 & 1.3 & 1.3 & 1.3 & 1.3 & 1.3 & 1.3 & 1.3 \\
& 155.6 keV$\rightarrow$ g.s. & 1.1 & 1.1 & 1.2 & 1.2 & 1.2 & 1.2 & 1.2 & 1.2 & 1.2 & 1.2 & 1.2 & 1.2 & 1.2 \\
\hline
\hline
& $ne_{26}$=30 & 0.05 & 0.1 & 0.2 & 0.3 & 0.4 & 0.5 & 0.6 & 0.7 & 0.8 & 0.9 & 1.0 & 1.1 & 1.2\\
\hline
& g.s. $\rightarrow$ g.s. & 0.9 & 1.5 & 1.9 & 2.0 & 2.0 & 2.0 & 2.0 & 2.0 & 2.0 & 2.0 & 2.1 & 2.1 & 2.1\\
& 87.2 keV$\rightarrow$ g.s. & 1.0 & 1.1 & 1.3 & 1.3 & 1.3 & 1.3 & 1.3 & 1.3 & 1.3 & 1.3 & 1.3 & 1.3 & 1.3 \\
& 155.6 keV$\rightarrow$ g.s. & 1.0 & 1.1 & 1.2 & 1.2 & 1.2 & 1.2 & 1.2 & 1.2 & 1.2 & 1.2 & 1.2 & 1.2 & 1.2 \\
\hline
\end{tabular}
\tablecomments{The values in the first line of each table are the temperature (in units: GK).
$ne_{26}$ is the electron number density in unit of 10$^{26}$ cm$^{-3}$.} 
\end{center}
\end{table}

\begin{deluxetable}{ccccc}
\tablecaption{Updated rates in network \label{tab:list} }
\tablehead{
\colhead{($n$,$\gamma$) Reaction} & \colhead{Comments}& \colhead{Reference} 
}
\startdata
$^{23}$Na($n$,$\gamma$)$^{24}$Na & Experiment & \cite{23Na} \\
$^{24}$Mg($n$,$\gamma$)$^{25}$Mg & Experiment & \cite{Mg} \\
$^{25}$Mg($n$,$\gamma$)$^{26}$Mg & Experiment & \cite{Mg} \\
$^{26}$Mg($n$,$\gamma$)$^{27}$Mg & Experiment & \cite{Mg} \\
$^{54}$Fe($n$,$\gamma$)$^{55}$Fe & Experiment & \cite{54Fe} \\
$^{56}$Fe($n$,$\gamma$)$^{57}$Fe & Experiment & \cite{Fe} \\
$^{57}$Fe($n$,$\gamma$)$^{58}$Fe & Experiment & \cite{Fe} \\
$^{59}$Fe($n$,$\gamma$)$^{60}$Fe & Experiment & \cite{59Fe} \\
$^{58}$Ni($n$,$\gamma$)$^{59}$Ni & Experiment & \cite{58Ni} \\
$^{62}$Ni($n$,$\gamma$)$^{63}$Ni & Experiment & \cite{6263nitof} \\
$^{63}$Ni($n$,$\gamma$)$^{64}$Ni & Experiment & \cite{63nidance} \\
$^{63}$Cu($n$,$\gamma$)$^{64}$Cu & Experiment & \cite{63Cu} \\
$^{64}$Zn($n$,$\gamma$)$^{65}$Zn & Experiment & \cite{Zn} \\
$^{70}$Zn($n$,$\gamma$)$^{71}$Zn & Experiment & \cite{Zn} \\
$^{70}$Ge($n$,$\gamma$)$^{71}$Ge & Experiment & \cite{70ge} \\
$^{72}$Ge($n$,$\gamma$)$^{73}$Ge & Experiment & \cite{72ge} \\
$^{73}$Ge($n$,$\gamma$)$^{74}$Ge & Experiment & \cite{73Ge} \\
$^{92}$Zr($n$,$\gamma$)$^{93}$Zr & Experiment & \cite{92Zr} \\
$^{93}$Zr($n$,$\gamma$)$^{94}$Zr & Experiment & \cite{93Zr} \\
$^{94}$Zr($n$,$\gamma$)$^{95}$Zr & Experiment & \cite{94Zr} \\
$^{95}$Zr($n$,$\gamma$)$^{96}$Zr & Experiment & \cite{95Zr} \\
$^{96}$Zr($n$,$\gamma$)$^{97}$Zr & Experiment & \cite{96Zr} \\
$^{147}$Pm($n$,$\gamma$)$^{148}$Pm & Experiment & \cite{147Pm} \\
$^{154}$Gd($n$,$\gamma$)$^{155}$Gd & Experiment & \cite{154gd} \\
$^{197}$Au($n$,$\gamma$)$^{198}$Au & Experiment & \cite{197au} \\
$^{16}$O($n$,$\gamma$)$^{17}$O & Theory & \cite{He_2020} \\
$^{17}$O($n$,$\gamma$)$^{18}$O & Theory & \cite{17O} \\
$^{13}$C($\alpha$,n)$^{16}$O & Experiment & \cite{13C} \\
$^{59}$Fe($\beta^{-}\bar{\nu}$)$^{59}$Co & Experiment & \cite{59Febeta} \\
\enddata
\end{deluxetable}

\bibliography{sample631}{}
\bibliographystyle{aasjournal}



\end{document}